\documentclass[11pt]{amsart}
\usepackage{graphicx}
\def\semicolon{\nobreak\mskip2mu\mathpunct{}\nonscript\mkern-\thinmuskip{;}
\mskip6muplus1mu\relax} % This defines the semicolon command
\usepackage[colorlinks = true,
            linkcolor = red,
            urlcolor  = magenta,
            citecolor = black,
            anchorcolor = blue]{hyperref}
\usepackage{color}
\usepackage[alphabetic]{amsrefs}
\usepackage{tikz}
\usetikzlibrary{arrows}
\usepackage[all]{xy}
\usepackage{wrapfig}
\usepackage{geometry}                % See geometry.pdf to learn the layout options. There are lots.
\usepackage{amssymb}
\usepackage{epstopdf}
\usepackage{mathrsfs}
\usepackage[mathcal]{eucal}

\newcommand{\rmb}{\mathbb{R}}

\newcommand{\Kr}[1]{\mathcal{#1}}
\newcommand{\po}[1]{\frac{\partial}{\partial #1}}
\newcommand{\dpo}[1]{\frac{\delta}{\delta #1}}

\newcommand{\dpod}[1]{\partial_{#1}}

\newcommand{\ppFp}[1]{\psi^{i}\psi^{j}F_{ij~a}^{b}\pi_{b}#1}

\newcommand{\corr}[1]{\langle #1 \rangle}
\newcommand{\bigcorr}[1]{\big\langle #1 \big\rangle}
\newcommand{\Bigcorr}[1]{\Big\langle #1 \Big\rangle}
\newcommand{\myref}[1]{~{(\ref{#1})}}

\newcommand{\massey}{\mathop{\mathrm{MP}}}

\newcommand{\beq}{\begin{equation}}
\newcommand{\eeq}{\end{equation}}

\newcommand{\mo}{\mathcal{O}}
\newcommand{\mon}{\mathcal{O}^{(1)}}
\newcommand{\mbr}{\mathbb{R}}

\newtheorem{thm}{Theorem}
\newtheorem{define}{Definition}
\newtheorem{conj}{Conjecture}

\makeindex

\begin{document}

\title[Morse Theory in Field Theory]
{MORSE THEORY IN FIELD THEORY\footnotemark}
\author{Peter Koroteev} 
\author{Andrey Zayakin}

\address{Peter Koroteev\newline 
\newline
Institute for Theoretical and Experimental Physics\newline
117259 Moscow, Russia\\}

\address{Moscow Institute of Physics and Technology\newline
141701 Dolgoprudny, Moscow region, Russia\\}

\address{Institute for Nuclear Research\newline
142190 Moscow, Russia\newline}

\address{Andrey Zayakin\newline
\newline
Institute for Theoretical and Experimental Physics\newline
117259 Moscow, Russia\\}

\address{Moscow State University\newline
119991 Moscow, Russia\\}

\address{Bogoliubov Laboratory of Theoretical Physics\newline
141980 Dubna,
Moscow region, Russia}

\begin{abstract}
We describe correlations functions of topological quantum
mechanics (TQM) in terms of Morse theory. We review the basics of topological field theories
and discuss geometric and algebraic interpretations of TQM. We prove that correlators in TQM
can be expressed via intersection numbers of certain submanifolds of
the target space with paths of steepest descent between critical
points of a Morse function. In the end we conjecture another correspondence between quantum mechanics correlators and integrals of Massey products of certain cohomology classes.
\end{abstract}

\maketitle

\section*{Introduction}
Topological quantum field theories and topological sting
theories originating from the works of E. Witten and others
\cite{WittenType,WittenTop,Witten} proved to be extremely helpful in understanding of
important mathematical problems.

Main feature of topological field theories (TFT) is independence of
correlation functions on metric and coordinates
\cite{Birmingham:1991ty}. In TFTs there are no propagating (local)
degrees of freedom,  vacuum expectation values of operators and
transition amplitudes (both further referred to as
``correlators'') in them are dependent only on topology of the
target manifold.

In this paper we employ for our purposes a simple example of TFT
-- Topological quantum mechanics (TQM) with a BRST-like invariant
action. It was shown \cite{Cohen,Labastida} that in
zero--dimensional analog of this theory partition function is
equal to Euler character of the target manifold.

We have two main aims in this paper: the first is to make manifest the
correspondence between TFT and geometry of target manifold; the
second aim is to study the correspondence between TFT and a
differential graded algebra of cohomology classes on target
manifold. The first aim is reached by us within a proof of a
reasonable (``physical'') level of strictness, whereas the second
is only conjectured and studied phenomenologically.

First we propose a geometrical interpretation of TQM developed in
\cite{Lysov,Losev:2005ze}. We prove that there is a correspondence
between a special kind observables and 1 codimensional cycles on
target manifold. Moreover, transition amplitudes in the theory
correspond to intersection indices of paths of steepest descent
and cycles. This correspondence is proven using path integral
representation of correlation function. Establishing a
correspondence between TQM and topology of the target manifold we
find a geometrical interpretation of all quantities in the theory.
It is also shown that correlator can be introduced independently
as an integral of pull--backs of forms corresponding to
observables over moduli space of graph embeddings into target
manifold.

In \cite{Lysov} it was shown that correlators in TQM satisfy the
so-called anticommutativity equation which is a general property
of TQM. This allows us to conclude that the same equation holds
for the intersection numbers. Thus an interesting mathematical fact is
proven by ``physical means''.

Second, we formulate a conjecture relating correlators in TQM with
algebraic operation -- Massey product -- on cohomology classes of the target manifold. 
The conjecture, together with the previous property of
correlators, makes it possible to relate Massey products and
the intersection numbers.

\section{Topological Quantum Mechanics. Overview}
\subsection{The Setup}
Quantum Mechanics can be considered as the simplest version of
Topological Field Theory (TFT). TQM is based on the following set
of axioms
\begin{enumerate}
\item Hamiltonian $H\in\mbox{End}_{\mathbb{R}}(\mathcal{H})$
acting on a Hilbert space of states
$\mathcal{H}=\mathcal{H}_0\oplus\mathcal{H}_1$ can be represented
as
$$
H = [Q,G],\,\,$$ where
$Q,G\in\mbox{End}_{\mathbb{R}}(\mathcal{H})$ are odd nilpotent
operators $$Q^2= 0, \,\,G^2 = 0,\,\,$$ brackets stand for
supercommutator, i.e. for two operators $A,B$ with parities $a,b$
respectively one has
$$
[A,B] = AB - (-1)^{ab}BA
$$
Hamiltonian annihilates vacua space
$$H\mathcal{H}_0 = 0, $$  which is postulated to be
non-empty. $H$ is positively defined on $\mathcal{H}_1$, commutes
with parity operator $(-1)^F$, whereas $Q$ anticommutes
$$
H(-1)^F = (-1)^F H,\,\,\,\, Q(-1)^F = -(-1)^F Q
$$
Here $F$ is a fermion number. \item Observables
$\Kr{O}_i\in\mbox{End}_{\mbr}\Kr{H}$ in TQM form the following
algebra
\begin{equation}\label{algebra}
\Kr{O}_i\Kr{O}_j = C^{k}_{ij}\Kr{O}_k,
\end{equation}
where $C^{k}_{ij}$ are its structure constants.
\end{enumerate}
\textbf{Important property of TQM.} In all TFTs correlation
functions are independent on
coordinates\cite{Birmingham:1991ty,Losev:2005ze}. But this
independence should be treated carefully. In the above setup this
property is valid if~\footnote{Operators $\Kr{O}$ satisfying these
equations are referred to as zero-observables.}
\begin{equation}
[Q,\Kr{O}_i] = 0
\end{equation}
then each correlator
\begin{equation}
\bigcorr{\Kr{O}_{i_1}(t_1)\dots\Kr{O}_{i_m}(t_m)} =
\mbox{Tr}(-1)^F e^{-t_1 H}\Kr{O}_{i_1}e^{(t_1-t_2)H}\dots
e^{(t_{m-1}-t_m)H}\Kr{O}_{i_m}e^{t_m H},
\end{equation}
where trace is taken over the Hilbert space $\Kr{H}$ is
independent on coordinates. But correlator may jump after
interchanging of some observables, so their order should be
preserved in the consideration.

\subsection{Deformation and One-Observables}
Let us deform operator $Q$ as $$Q\to Q+\sum T^A\Kr{O}_A=Q+\Kr{O}$$
where $T_A$ are parameters(coupling constants), $\Kr{O}_A$ are
zero-observables. Then the Hamiltonian becomes
\begin{equation}\label{deform}
H=[Q,G]\to[Q,G]+[\Kr{O},G]=H_0+H_1
\end{equation}
Considering $H_1$ as interaction Hamiltonian we can rewrite
evolution operator and the derivative of 1-point correlator
$\Kr{O}_1(t)$
\begin{equation}\label{1pointcorr}
\po{T^A}\corr{\Kr{O}_1(t_1)}=-\int\limits_{0}^{t_1}
d\tau\,\mbox{Tr}(-1)^F e^{-(t_1-\tau)H}[\Kr{O}_A,G]e^{-\tau
H}\Kr{O}_1 =
\int\limits_{0}^{t_1}\corr{\Kr{O}^{(1)}_A(\tau)\Kr{O}_1(0)}d\tau,
\end{equation}
where $\Kr{O}_A^{(1)}(t)\equiv - [\Kr{O}_A(t),G]dt$ is referred to
as 1-observable. One-observable is an 1-form on $\mbr$.

\subsection{Generating Function for Correlators}
The following property holds for a correlator\cite{Lysov}
\begin{equation}\label{exp}
\corr{\Kr{O}_{A_i}}^{A}_{B\,deformed} =
\po{T^{A_i}}\corr{\Kr{O}}^A_{B\,deformed} = \corr{\Kr{O}_{A_i}
e^{\int_{\mbr}[\Kr{O}(t),G]dt}}^A_B.
\end{equation}
Here $\bigcorr{..}_{deformed}$ denotes vacuum expectation value in
an interacting (deformed) theory, $\bigcorr{..}$ -- the same
quantity in a free (non-deformed) theory.

We can expand the exponent in (\ref{exp}) in the following Taylor
series according to parameters $T_A$
\begin{equation}\label{expect}
\Kr{F}^A_B(T) \equiv \corr{T\big\{\Kr{O}_{A_1}
e^{\int_{\mbr}[\Kr{O}(t),G]dt}\big\}}^A_B=
\sum\limits_{m=1}^{\infty}\Kr{F}^{A}_{B;\,A_1 \dots,
A_{m}}T^{A_2}\dots T^{A_{m}},
\end{equation}
where the coefficient is expressed via
\begin{align}\label{F}
&\Kr{F}^{A}_{B;\,A_1\dots A_m} \equiv
\frac{1}{(m-1)!}\Bigcorr{\mbox{T}\Big\{\,\mo_{A_1}
\Big[\prod\limits_{i=2}^{m}\int\limits_{\rmb}[\mo_{A_i}(t_i),G]dt_i\Big]\Big\}}^{A}_{B}\notag\\
&= \int\limits_{\mathbb{R}_{+}^{m-1}}d\tau^1\dots
d\tau^{m-1}\bigcorr{\mo_{A_1} G e^{-\tau_1 H}\mo_{A_2}(0)G
e^{-\tau_2 H}\dots \mo_{A_m}(0)}^{A}_{B}
\end{align}
Parameters $T_A$ have the meaning of coupling constants here and
$T\{..\}$ stands for chronological ordering. The whole expression
(\ref{expect}), if interpreted physically, corresponds to the
vacuum expectation value of $\mo_{A_1}$ in the theory with
interaction $H_1$. If all $\Kr{O}_{A_i}=\Kr{O}$ are the same, the
operator $K=\int_{0}^{+\infty}G\exp(-H\tau)d\tau$ being
introduced, the above formula can be compactly rewritten as
\begin{equation}
\Kr{F}^{(m)\,A}_{\,\,B}=\bigcorr{\Kr{O} K \Kr{O}\dots
K\Kr{O}}^A_B,
\end{equation}
where $\Kr{F}^{(m)}$ is a short notation for the value determined
in (\ref{F}).

It was shown in \cite{Lysov} that the following anticommutativity
equation holds for correlators
\begin{equation}\label{anticommutativity}
\Kr{D}\Kr{F} + \frac{1}{2}[\Kr{F},\Kr{F}] = 0\,\,\,\,
\text{or}\,\, (\Kr{D}+\Kr{F})^2=0,
\end{equation}
where $\Kr{D}=C_{AB}^{K} T^A T^B \po{T^K}$ is BRST operator
(Chevaller differential). Here $\Kr{D}^2=0$ if $C_{AB}^{K}$ is
antisymmetric with respect to $A$ and $B$. (Vacuum indices are
omitted here). We shall make use of it later.

\section{Geometrical Interpretation of TQM}
\subsection{Path Integral Representation of TQM}
In this Section we use a theory which is a particular case of TQM.
We are going to proceed in a slightly unconventional way, namely,
first defining transition amplitudes and afterwards deriving the
action functional from them. This will be done to make the
geometrical interpretation of the theory more manifest.

Let $\Kr{M}$ be a smooth closed oriented Riemannian n-manifold,
$f$ is a Morse function on it and $v$ is gradient vector field
constructed by means of this Morse function, $\mbox{CP}(\Kr{M})$
the space of its critical points~\footnote{see next subsection for
details}. Let $A,B\in \mbox{CP}(\Kr{M})$ be critical points with
indices $p+1$ and $p,\,\,p=0,\dots,n-1$
 respectively, and $\Gamma^{A}_B$ be a gradient
curve initiating at $A$, terminating at $B$ and satisfying the
following set of ODEs
\begin{equation}\label{PSD}
\dot{x}^i = v^{i}.
\end{equation}
Its solutions are integral curves of $v$, or paths of steepest
descent (PSD). The worldsheet of the theory is a line $\rmb$,
targetspace is $\Kr{M}$ and embeddings $x\in\mbox{Map}^A_B$
\begin{equation}
\mbox{Map}^A_B = \{x(t)\in C^{\infty}(\mathbb{R}^1,\Kr{M}))
\,,\quad  x(-\infty)=A,\,\,x(+\infty)=B\},
\end{equation}
satisfying \eqref{PSD}
\begin{figure}[t]
\begin{center}\label{embedding}
\includegraphics[height = 4cm,width=6cm]{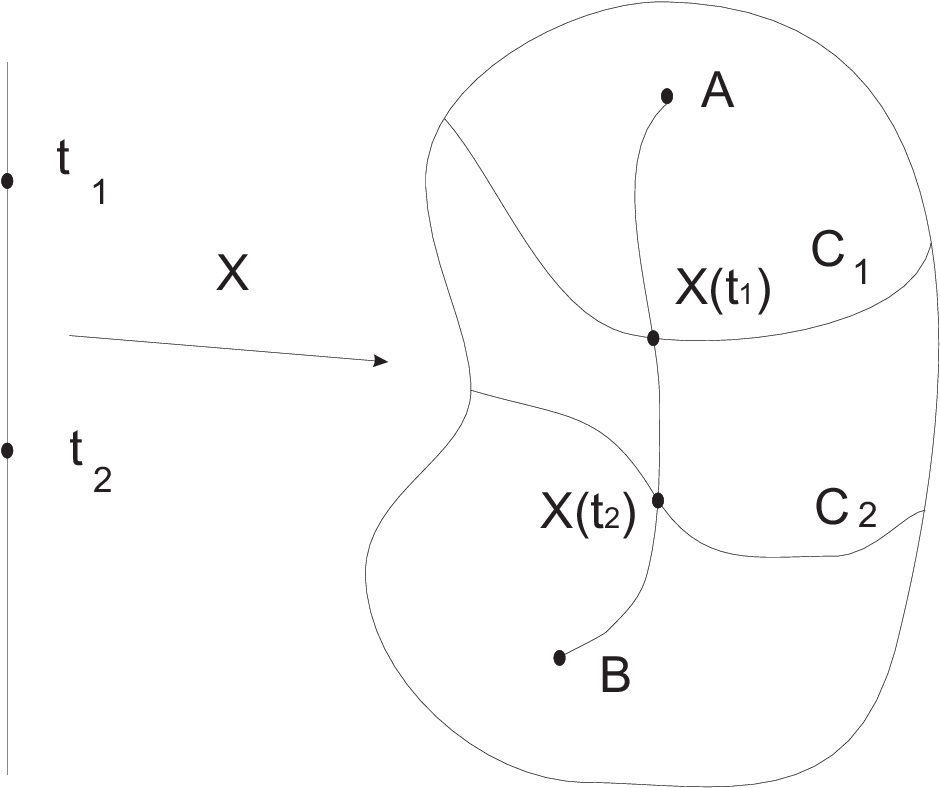}
\caption{Worldsheet and target space of the theory}
\end{center}
\end{figure}
So we embed a line into $\Kr{M}$ with fixed images of $\pm\infty$
and $x$ are local coordinates on the targetspace, requiring it to
be one of the rigid paths of steepest descent between $A$ and
$B$~(fig. \ref{embedding}). In further considerations we will
imply that the above boundary conditions are satisfied. As for the
dimension of the space $\mbox{Map}^A_B$ the following statement is
valid\cite{Fukaya1}
\begin{equation}
\alpha:= \mbox{dim}\,\mbox{Map}^A_B = \mbox{ind}A-\mbox{ind}B-1.
\end{equation}
For example, for $f$ being height function on manifold
$\Kr{M}=\mathbb{S}^1\subset\mathbb{R}^2$ we have $\alpha=0$, as it
is typical in most cases for $A$ and $B$ with indices different by
1. There is usually a finite number of paths between them.  This
explains the term ``rigid'': these paths of steepest descent
cannot be continuously deformed into other paths of steepest
descent.

To develop a quantum theory means to describe the states and
transition amplitudes among them. Transition amplitudes are given
by path integrals with appropriate boundary conditions. The key
point in understanding the geometrical essence of  TQM described
below is that transition amplitudes in it can be constructed form
a purely geometric object. Indeed, consider the following path
integral, which does not yet have anything to do with any TFT. At
this stage it is just a functional of vector fields on a manifolds
$\mbox{exp}(-S)$, where $S$ is an action is a delta-functional on
elements of $\mbox{Map}^A_B$ space -- space of paths of steepest
descent. So we have the following path integral
\begin{equation}\label{generating}
Z_{AB} = \int\Kr{D}x\,\delta[V]\int\limits_{-\infty}^{+\infty}
dt\, \mbox{det}(\nabla_i V^j(x(t))),
\end{equation}
where $x(t)$ takes its value in $\Kr{M}$, $\nabla_i$ is a
covariant derivative with Levi-Civita connection and $V=\dot{x}-v$
and $\delta[V]$ is a delta-functional. $V=0$ corresponds to path
of steepest descent. The determinant det is a finite-dimensional
determinant of matrix $\mbox{det}\,(\int dt\nabla_i V^j(x(t)))$.
The following boundary conditions $x(-\infty)=A,\,x(+\infty)=B$
are imposed\footnote{Analogous boundary conditions will be imposed
upon each path integral in this paper, unless specified
otherwise.}; we remind again that $A,B$ are critical points of the
Morse function on the manifold. The measure is here standard
Feynman's measure $\Kr{D}x=\Pi_{k=1}^{\infty}dx(t_k)$ with fixed
endpoints. By construction this integral counts \textbf{the number
of PSDs with signs}. This technique is an infinite dimensional
generalization of Mathai--Quillen method\cite{MQ}.

Below it will be made clear that there is a one-to-one
correspondence between critical points of Morse function on the
manifold and vacua in TQM; moreover, $Z_{AB}$ corresponds to an
transition  amplitude from vacuum $|A \rangle$ to vacuum
$|B\rangle$. The formula (\ref{generating}) can be rewritten as a
path integral over even functional variables $\Kr{D}x\Kr{D}p$ and
odd functional variables $\Kr{D}\psi\Kr{D}\pi$, measure
$\Kr{D}x\Kr{D}\psi$ being measure with fixed endpoints, measure
$\Kr{D}p\Kr{D}\pi$ being measure with arbitrary
endpoints.\footnote{Thus in the formula~\eqref{genfnc} there is
one more integral over $\Kr{D}p\Kr{D}\pi$ then over
$\Kr{D}x\Kr{D}\psi$}
\begin{equation}\label{genfnc}
Z_{AB}=\int\Kr{D}x\Kr{D}\psi\Kr{D}p\Kr{D}\pi\,e^{-S[x(t), p(t) ,
\psi(t),\pi(t)]},
\end{equation}
where the functional $S[x(t),p(t),\psi(t),\pi(t)]$ is naturally
identified with action and looks like
\begin{align}\label{action}
S = \int\limits_{-\infty}^{+\infty} &dt\,
\Big[p_{a}\big(\dot{x}^{a} - v^{a}\big) -
\psi^{i}\big(\nabla_{i}v^{a}(x)\big)\pi_a +
\epsilon\ppFp{\pi_{c}\eta^{ac}} + \epsilon
p_{a}p_{b}\eta^{ab}\Big]\notag\\ &=\int p_a dx^a + \pi_a d\psi^a -
[Q,G]dt=\int PdQ-Hdt,
\end{align}
for generalized super coordinates $Q$ and momenta $P$. $A$ is the
affine connection which is related with Levi-Civita connection as
$A^a_{ib}=e^a_k e^l_b\Gamma^k_{il}$ where $g_{ij}=\sum_a e^a_i
e^a_j$. Accordingly, $x(t),p(t),\psi(t),\pi(t)$ are identified
with dynamic variables, and $Z$ with transition amplitude,
$\epsilon$ plays a role of mass parameter; alternatively, it can
be thought of as coupling constant. In \eqref{action}
\begin{align}\label{qn}
Q &= \psi^{i}\nabla_{i} + p_{a}\dpo{\pi_{a}} +
\psi^{i}\psi^{j}F_{ij ~ b}^{~ a}\pi_{a}\dpo{p_{b}},\notag\\
G&=\pi_a v^a + \epsilon \pi_a p_b \eta^{ab},
\end{align}
where $F_{ij ~ b}^{~ a}=\dpod{i}A^a_{jb}+A^{a}_{ic}A^c_{jb}$ is a
curvature tensor.

Critical points $A,B$ are identified with vacua of the theory due
to the following reason: in Lagrangian formalism $V(x)\sim
(v^i(x))^2$ has the meaning of potential\footnote{This relation
becomes apparent after Gauss integration by $p$}, due to
positive-definiteness, its zeros are its minima; the zeros of a
gradient vector field are critical points of the corresponding
Morse function.

We provide the following Table of correspondence.
\begin{center}
{\renewcommand{\arraystretch}{0}%
\begin{tabular}{|c|c|c|}
\hline  Abstract TQM & PI representation of TQM& Morse theory \\
\hline  \rule{0pt}{1pt}&\\ \hline  Vacua space $\Kr{H}_0,\,\, |A
\rangle$& Minima $A$ of the potential& Critical points
$\mbox{CP}(f,\Kr{M}),\,\, A$\\ \hline  Observables $\Kr{O} $
&$\delta$-functions (1-forms) on cycles& Cycles $C\subset\Kr{M} $
\\ \hline  Amplitude $\langle A | B \rangle $&Amplitude $\langle A | B \rangle $ &
\# PSDs from A to B \\ \hline  Operator $Q$&
$\psi^{i}\nabla_{i} + p_{a}\dpo{\pi_{a}} + \psi^{i}\psi^{j}F_{ij ~
b}^{~ a}\pi_{a}\dpo{p_{b}}$& de-Rham diff. $d$
\\ \hline  Operator $G$&$\pi_a v^a + \epsilon \pi_a p_b \eta^{ab}$& Inner product $\,\iota_v $ \\
\hline
\end{tabular}
}
\end{center}

\subsection{\label{obser}{Morse Theory, Witten Complex and n-Matrices}}
In our theory the space $\Kr{H}_0$ of vacua corresponds to the
space of critical points of function $f$ on $\Kr{M}$
$\mbox{CP}(f,\Kr{M})$. So vacua $|A \rangle,|B\rangle$ correspond
to critical points $A,B$ and the transition between the initial
and the final state corresponds to motion of the point on manifold
from $A$ to $B$ by rigid path $\Gamma^A_B$.

Let $\mbox{CP}^i$ be the linear space of formal linear
combinations of all critical points of $\Kr{M}$ of index $i$. The
following complex of chains $\mbox{CP}^i$
\begin{equation}
  \dots \longrightarrow \mbox{CP}^{k-1} \longrightarrow
  \mbox{CP}^{k} \longrightarrow \mbox{CP}^{k+1} \longrightarrow \dots
\end{equation}
is said to be Witten complex \cite{Witten,Fukaya1}. Here $n$
($n^2=0$) is a coboundary operator which increases grading in the
complex is given by the explicit formula is
\begin{equation}\label{nviagamma}
n|B\rangle = \sum\limits_{\Gamma^A_B}\mbox{sign}\Gamma^A_B|A
\rangle,
\end{equation}
where $|B\rangle\in{\mbox{CP}^i},|A \rangle\in{\mbox{CP}^{i+1}}$
and $\mbox{sign}\Gamma^A_B$ in the formula (\ref{nviagamma}) gives
a sign from each PSD from $A$ to $B$ and it was defined in
Witten's paper \cite{Witten}. If there are several PSDs then their
signs are summed up so we can rewrite the formula in the following
more convenient matrix notation
\begin{equation}
n|B\rangle = \sum\limits_{A}n^A_B |A \rangle,
\end{equation}
where $n^A_B$ is a matrix element, which is equal to the number of
PSDs computed with signs, initiating at $A$ and terminating at
$B$.

Along with $n^A_B$ we can introduce $n^A_B(C_1,\dots,C_m)$ -- a
number of PSDs from $A$ to $B$ intersecting cycles
$C_1,\dots,C_m\subset\Kr{M}$. Here a transversal intersection of
1-dimensional cycles and curve is assumed. Also we consider each
cycle intersecting a PSD only once.

\begin{define}
Under the above assumptions,
\begin{equation}
n^A_B(C_1,\dots,C_m)=\sum
\limits_{\Gamma}\prod\limits_{i=1}^{m}\mbox{ind}(C_i,\Gamma^A_B),
\end{equation}
is said to be higher Morse differential (or n-matrices). Here
$\mbox{ind}(C_i,\Gamma^A_B)$ is an intersection index of the
objects into parentheses. \end{define}

Eventually we have a family of operators, represented by matrices
$n^A_B,\,\,
n^A_B(C_1,\dots,C_m)\\\in\mbox{Hom}(\mbox{CP}^i,\mbox{CP}^{i+1})$
and as all these objects are nilpotent one can consider a complex
for each operator $n^A_B(C_1,\dots,C_m)$ analogous to Witten
complex for $n^A_B$.

\subsection{\label{geom}Correlator via Intersection Numbers}
Here we are going to introduce an explicit formula for correlator
in Morse theory version of TQM. It will be introduced as a
definition but in the next subsection we'll show that the given
expression really can be expressed via path integral.

First we take an embedding
\begin{equation}
x\in\mbox{Map}^A_B:=\{x\in C^{\infty}(\mbr,\Kr{M})\semicolon
x(-\infty)=A, x(+\infty)=B\}
\end{equation}
for $A$ and $B$ as critical points of certain Morse function and
obtain images of points $t_1,\dots,t_m$ by this map
$x(t_1),\dots,x(t_m)$. One can treat an evaluation map
\begin{align}
ev: \mathbb{R}^1\times \mbox{Map}^A_B &\longrightarrow
\Kr{M}\notag\\
 (x,t) &\longmapsto x(t)\notag
\end{align}
We will employ a pull-back of differential forms
\begin{equation}
ev^{\ast}: \Omega^{\bullet}(\Kr{M}) \longrightarrow
\Omega^{\bullet}(\mathbb{R}^1\times\mbox{Map}^A_B )
\end{equation}
For some form $\omega\in\Omega^{\bullet}(\Kr{M})$ this map is the
following\footnote{I don't know how $d\varphi$ can be described
rigourously but its particular form won't be important for us}
\begin{equation}
ev^{\ast}\omega_I(x)dx^I=\omega_I(x(t))(\dot{x}^I
(t)dt+d\varphi^I),
\end{equation}
where differentials $d\varphi^I$ belong to $\mbox{Map}^A_B$ space.

We construct transversal cycles\footnote{thus each cycle has only
one common point with AB} to path $\mbox{AB}$ $C_1,\dots,C_m$ on
$\Kr{M}$ so that $x(t_i)\in C_i, \,\, \mbox{codim}C_i = 1$. Our
next step here is to build closed differential form on $\Kr{M}$
which is delta-function on a cycle

\begin{define} Form $\omega(x)$ is said to be a delta-form on
cycle $C$ and denoted
\begin{equation}
\omega(x) = \delta_{C} \notag
\end{equation}
if
\begin{equation}\label{delta1}
\int\limits_{\Kr{M}}\omega\wedge\delta_{C} = \int\limits_C\omega
\end{equation}
\end{define}

And it is a pull--back of delta form which is referred to as
observable in TQM.
\begin{equation}
\Kr{O} = ev^{\ast}\omega_i. \notag
\end{equation}
We introduce a new space $\mathfrak{M}\equiv\mbox{Map}^A_B
\times\mbr^1$. Then the following integral is considered
\begin{equation}\label{corr}
\int\limits_{\mathfrak{M}\times\mathbb{R}^{m-1}_{+}}ev^{\ast}\delta_{i_1}\wedge\dots\wedge
ev^{\ast}\delta_{i_m}.
\end{equation}
Here $\mathbb{R}^{m-1}_{+}$ is a moduli space of embeddings of
graphs $t_1$---$t_2$---\dots
---$t_m$ into all paths of steepest descent between
$A$ and $B$. The substantial statement is that the above integral
is equal to $n^A_B(C_1\dots C_m)$.

Indeed, as Map in our case is zero dimensional and represents a
finite set of points and
$\mathfrak{M}\times\mathbb{R}^{m-1}_{+}=\mbox{Map}^A_B\times
\bigsqcup\mathbb{R}^{m-1}_{+}\times\mbr^1$ the integral
(\ref{corr}) is equal to the sum of integrals corresponding to
each PSD in Map space. Then each term in this sum equals
$\prod\mbox{ind}(C_i,\Gamma)$ by construction. So one has that the
integral (\ref{corr}) is equal to $n^A_B(C\dots C)$. In fact the
integrand after integrating over Map equals wedge product of delta
forms on cycles and delta form on PSD. As all cycles are
transversal to the path the dimension of the set in the
intersection can be equal only to 0. If there is no intersection
of cycle with PSD the answer is 0 for the whole integral and
equals 1 if each cycle have an intersection with the curve. In
fact the integral under consideration is a sum of $\#\mbox{Map}$
integrals over m-dimensional manifold of m-dimensional
delta-function.

\subsection{Correlator via Path Integral}
Now we will convince the reader that $\Kr{F}^{A\,(m)}_{B}$ in
\myref{F} is equal to $n^A_B(\underbrace{C\dots C}_{m})$ provided
an appropriate correspondence for the abstract operator
$G\Kr{O}_{A_i}$ is specified in path integral formulation.

\begin{thm}
Let $v$ be a smooth vector field on $\Kr{M}$ and $\mo^{(1)}_{A_i}$
be 1-observable in TQM. Let $G$ be the inner product $G=\iota_v$. Then the following equality between
correlator and intersection numbers holds
\begin{equation}
\Kr{F}^A_{B;\,A_1\dots A_m}=n^A_B(C_1,\dots,C_m),
\end{equation}
\rm

\textbf{Proof.} One
can see that the following correspondence arises from the theorem
\begin{equation}
G\Kr{O}_{A_i}=\delta(x^n(0)-x^n(t_i))v^n(x(t_i)),
\end{equation}
Here local coordinates are chosen in such a way that vector field
$v$ on $\Kr{M}$ in the vicinities of intersection points with path
of steepest descent has the only one nonzero component $v^n$,
$x^n$ is the n-th component of coordinate $x$,
 and $x(t_i(\tau))$ are images of points
$t_1,\dots,t_m$ dependent on $\tau$ by embedding $x$.\footnote{by
$\tau$ we imply $\tau_2,\dots,\tau_m$} One can see that after
applying inner product $\iota_v$ to one
observable a terrible differential $d\varphi$ vanishes.

Representation of the correlator~\eqref{F} via path integral
yields
\begin{equation}
\Kr{F}^{(m)}=\int\limits_{\mathbb{R}_{+}^{m-1}}d\tau_2\dots
d\tau_m\int e^{-S}\mo_{A_1} \mon_{A_2}\dots\mon_{A_m}.
\end{equation}
Here $\tau_i=|t_i-t_{i-1}|,\,\,i=\overline{2,m}$ parameterize the
moduli space of embeddings of $\mbr$ with marked points
$t_1,\dots,t_m$ into target manifold $\Kr{M}$ (see Subsections
\ref{obser} and \ref{geom}).

Then the correlator expansion coefficient can be expressed via
path integral
\begin{align}
\Kr{F}^{(m)\,A}_B&=\displaystyle\int d\tau_2\cdots d\tau_m
\int\Kr{D}x\Kr{D}p\Kr{D}\psi\Kr{D}\pi\exp
\left[-\int_{-\infty}^{\infty}dt\big(p_a(v^a(x)-\dot{x}^a)+\pi_a\dot{
\psi^a } -\psi^i(\nabla_i v^a)\pi_a\big)\right]\notag\\& \times
\delta(x^n(0)-x^n(t_1(\tau)))\prod\limits_{i=2}^{m}\delta(x^{n}(0)
-x^{n}(t_i(\tau)))\,v^{i}(x^n(t_i(\tau))),
\end{align}
As the operators contain no dependence on Grassman fields
$\psi(t),\pi(t)$, one can integrate them out, resulting in
$\det(\partial_\tau\delta_{i}^j-\partial_i v^j)$ in the numerator.
The integral over $p(t)$ can also easily be done, simply by the
definition of delta-functional. Therefore,
\begin{align}\label{evF}
\Kr{F}^{(m)\,A}_B&=\int d\tau_2\cdots d\tau_m\int\Kr{D}x
\det(\partial_t \delta_{i}^j-\partial_i
v^j)\,\delta[\dot{x}(t)-v(x)]\notag\\\times&\delta(x^n(0)
-x^n(t_1(\tau)))\prod\limits_{i=2}^{m}
\delta(x^n(0)-x^n(t_i(\tau)))v^n(x(t_i(\tau))).
\end{align}
One can take the integral over $\Kr{D}x$ away by virtue of the
delta-functional, replacing $x(t)$ by $\textsc{x}(t)$ -- solution
of classical Lagrange-Euler equations~\eqref{PSD}. However,
special cares should be taken due to the presence of zero modes in
these solutions. Hence, an integral over the space of collective
coordinates $\lambda$ remains after integrating the
infinite-dimensional $\Kr{D}x$ integral~\cite{Birmingham:1991ty}.
Geometrically $\lambda$ corresponds to shift of all points
$t_1,\dots,t_m$ along $\mbr$ keeping distances between each other
constant and parameterizes second multiplier in the definition of
$\mathfrak{M}$ (see \ref{geom}). Reexpressing the delta-functional
\begin{equation}
\delta[\dot{x}^i(0)-v^i(x)]=\sum_\Gamma \frac{\delta
[x^i(0)-\textsc{x}(t)]}{\vert\det( \dpod{t}\delta_k^j -
\dpod{k}v^j)\vert},\notag
\end{equation}
one cancels determinants (as it should be in a supersymmetric
theory) up to sign $(-)^a = \mbox{sign}(\det( \dpod{t}\delta_k^j -
\dpod{k}v^j))$ and obtains
\begin{align}
\Kr{F}^{(m)\,A}_B&=\sum_\Gamma\int d\tau_2\cdots d\tau_m\, \int
d\lambda\,(-)^a\,\delta(x^{(cl)\,n}_{\Gamma}(0)-\textsc{x}(t_1(\tau),\lambda))\notag\\
&\prod\limits_{i=2}^{m}
\delta(\textsc{x}(0)-\textsc{x}(t_i(\tau),\lambda))v^n(\textsc{x}(t_i(\tau)))
\end{align}
Integral over $\lambda$ plays a crucial role here. It allows us to
integrate out all the delta-functions, so that a regular
expression remains. The latter integral possesses structure
absolutely similar to that of the integral\myref{corr}, which was
obtained within a purely geometric construction of section
\eqref{geom}. Indeed, sum over $\Gamma$ and the integral over zero
mode $\lambda$ are equivalent to integration over $\mathfrak{M}$,
whereas the integrals over $\tau_i$ are taken over same manifolds
$\mathbb{R}^{m-1}_+$. One can make sure that the following
integral
\begin{equation}\label{unit}
\int d\tau_i\,(-)^a\,
v^n(\textsc{x}(t_i(\tau)))\,\delta(\textsc{x}(0)-\textsc{x}
(t_i(\tau)))=\mathrm{ind}(\Gamma,C_i)
\end{equation}
is an intersection index between $\Gamma$ and $C_i$. Therefore,
one comes using~(\ref{unit}) to the following expression
\begin{equation}\label{}
\Kr{F}^{(m)\,A}_B=\sum_\Gamma\prod\limits_{i=1}^{m}\mathrm{ind}
(\Gamma, C_i)=n^A_B(C\dots C).
\end{equation}
Trivially generalizing this result, we thus have proven that for
an arbitrary number of cycles,
\begin{equation}\label{correspondence}
\Kr{F}^A_{B;\,A_1\dots A_m}=n^A_B(C_1,\dots,C_m)
\end{equation}
$\Box$
\end{thm}

\subsection{Generating function for N-matrices}
The correspondence described in the above subsection is very
useful and has interesting consequences. Indeed as
$\Kr{F}^{(m)}=n^A_B(C_1\dots C_m)$ one can rewrite for
intersection numbers all relations valid for the correlators as
well. First we mean the anticommutativity equation. As before we
construct a generating function namely the whole matrix of them
\begin{align}
N^A_B(T)&:= n^A_B + n^A_B(C_i)T^i+ n^A_B(C_j,C_k)T^j
T^k+\dots\notag\\&+n^A_B(C_p,\dots ,C_q)T^p\dots
T^q+\dots=\sum\limits_{k}n^A_B(C^{(k)})T^{(k)},
\end{align}
where $t$ is a parameter. Actually several nonequivalent cycles
are admitted so one needs introducing the same number of
parameters. The above construction is an element of the space
$\mbox{Mat}_{N\times N}\otimes \mbr[T^1\dots T^l]$. Here $l$ is a
number of nonequivalent cycles on the manifold.

As anticommutativity equation holds for $N$ ($\Kr{D}$ is taken
from (\ref{anticommutativity}))
\begin{equation}
[\Kr{D}+N,\Kr{D}+N]=0\,\,\,\,\,\text{or}\,\,\,\,\Kr{D}N+\frac{1}{2}[N,N]=0
\end{equation}
we obtain interesting relations on intersection numbers in every
order in $T$. These equations are indeed very interesting
relations in the intersection theory.

\section{Algebraic interpretation of TQM}
Massey product is defined as follows
\begin{define}
Let $\alpha\in H^p(\Kr{M}),\,\beta\in H^q(\Kr{M}),\,\gamma\in
H^r(\Kr{M})$ and $\alpha\beta = 0,\,\beta\gamma = 0$. Then Massey
product $\mbox{MP}(\alpha,\beta,\gamma)$ is an element of the
following quotient space
\begin{equation}
H^{p+q+r-1}(\Kr{M})/[\alpha\smile
H^{q+r-1}(\Kr{M})+H^{p+q-1}(\Kr{M})\smile\gamma]
\end{equation}
Let cocycles $a,b,c$ be representatives of $\alpha,\beta,\gamma$
and cochains $u,v$ such that $du=ab$ and $dv=bc$. Then the cochain
$-uc+(-1)^p av$ is a cocycle and its cohomological class
represents $\mbox{MP}(\alpha,\beta,\gamma)$.
\end{define}

Higher Massey products are defined inductively via products of
less order Massey products. However, for higher order Massey
products to exist it is necessary for the lesser order Massey
products to be trivial. Massey product enables us to determine
homotopic class of the manifold up to torsion group.

One can see that the above construction is not well defined.
Nevertheless, this problem can be solved introducing the so-called
modified Massey product. We need $\alpha\beta,\,\beta\gamma$
vanished in cohomologies. If they are nonzero the above definition
fails. So we introduce the following operator
\begin{equation}\label{kdef}
K = d^{-1}\circ (id-\mbox{Pr}_{H}),
\end{equation}
where $\mbox{Pr}_{H}$ is a projection operator on de-Rham
cohomologies\footnote{Cohomologies here are treated as vector
space which}. So for each form $\omega$ one has
$(id-\mbox{Pr}_{H})\omega = d\chi$ being exact, the operator
$d^{-1}$ is well defined and $K\omega=\chi$. But another problem
arises here. Form $\chi$ is not closed any more.

\subsection{Conjecture}
As there is an embedding of the space of critical points of the
Morse function $\mbox{CP}^{\bullet}\hookrightarrow
H^{\bullet}(\Kr{M})$ in de Rham cohomology groups of $\Kr{M}$ (as
liner spaces) one can make the following
\begin{conj}
Let $A\in\mbox{CP}^{p+1}$ and $B\in\mbox{CP}^{p}$ be critical
points of indices $p+1$ and $p$ respectively and $|A \rangle\in
\Kr{H}_0^{p+1},\, |B\rangle\in\Kr{H}_0^p$ be representatives of
vacua space.

There is a correspondence between observables in TQM and forms in
de Rham cohomology groups of $\Kr{M}$
\begin{equation}
End(\Kr{H}_0)\ni\Kr{O}_{A_i}\longleftrightarrow\omega_{i}\in
H^{1}(\Kr{M})
\end{equation}
such that
\begin{equation}\label{conjecture}
\int\limits_{\mathbb{R}^{m-1}_{+}} d^{m}\tau
\,\Bigcorr{\Kr{O}_{\{A_{1}}Ge^{-\tau_2 H}\Kr{O}_{A_2}\dots
Ge^{-\tau_{m}H}\Kr{O}_{A_{m}\}}}^A_{B}=
\int\limits_{\Kr{M}}\tilde{\omega}^{A}\wedge\massey(\omega_1,\dots,\omega_m;\omega_B),
\end{equation}
where tilde stands for Poincar\'e duality.
\end{conj}
So if the conjecture is valid then we have an equality of three
objects of very different nature -- correlator in TQM,
intersection matrix $n^A_B(C_1,\dots, C_m)$ and the expression in
the r.h.s. of the above formula.

\section{Conclusion}
We have shown the geometrical pattern of TQM -- a toy model of
TFT. We expressed correlators via intersection numbers on target
manifold and made a conjecture that they can be expressed via
integral of Massey product. Now the main problem is to prove this
statement.

Quantum mechanics in the setup described tn this paper is a string
theory with string length equal to zero. Generalization of this
theory to topological sigma--model is a very interesting problem
to be solved.

\subsection*{Acknowledgements}
The authors are grateful to A. Gorsky, D. Krotov, D.
Volin, A. Chervov, M. Libanov, A. Litvinov and especially A. Losev
for fruitful discussions. This work is supported in part by grant
RFBR 01-02-17227.

\bibliographystyle{ieeetr}

\begin{thebibliography}{99}

\bibitem{Birmingham:1991ty}
  D.~Birmingham, M.~Blau, M.~Rakowski and G.~Thompson,
 % ``Topological field theory,''
  Phys.\ Rept.\  {\bf 209} (1991) 129.

\bibitem{Cohen} L.~Cohen and P.~Norbury, Morse Field Theory.
math.GT/0509631.

\bibitem{Fukaya1} K.~Fukaya. Morse homotopy and its quantization.
AMS/IP Studies in Advanced Math. 2, (1997) pp. 409 - 440

\bibitem{Fukaya2} K.~Fukaya. Morse homotopy, $A^{\infty}$
category and Floer homologies. Proc. of Garc  Workshop in
Geometry, Seoul National Univ.

\bibitem{Labastida} J.~Labastida. Morse Theory Interpretation
of Topological Quantum Field Theories. Commun. Math. Phys. 123,
641-658 (1989)

\bibitem{Losev:2005ze}
  A.~Losev and I.~Polyubin,
  %``Topological quantum mechanics for physicists,''
  JETP Lett.\  {\bf 82} (2005) 335.

\bibitem{Lysov}
  V.~Lysov,
  %``Anticommutativity equation in topological quantum mechanics,''
  JETP Lett.\  {\bf 76}, 724 (2002)
  [Pisma Zh.\ Eksp.\ Teor.\ Fiz.\  {\bf 76}, 855 (2002)]
  [arXiv:hep-th/0212005].

\bibitem{MQ} V.~Mathai and D.~Quillen. Superconnections, Thom classes and equivariant
differential forms. Topology 25:85-110, (1986).

\bibitem{WittenType} E.~Witten. Topological Quantum Field Theory.
Comm. Math. Phys. 117, 353-386 (1988).

\bibitem{WittenTop} E.~Witten, Topological sigma model.

\bibitem{Witten} E.~Witten, Supersymmetry and Morse Theory.
J.Diff.Geom. \textbf{17} (1982) 661.

\end{thebibliography}

\end{document}